\documentclass[conference]{IEEEtran}
\usepackage{amsmath}
\usepackage{amssymb}
\usepackage{amsthm}

\usepackage{graphicx}
\usepackage{cite}

\ifCLASSINFOpdf
\else
\fi

\begin{document}
\title{Affine Frequency 
Division Multiplexing with Subcarrier Power-Level Index Modulation for Integrated Sensing and Communications}

\author{\IEEEauthorblockN{Murat Temiz}
\IEEEauthorblockA{\textit{Department of Electronic and Electrical Engineering} \\
\textit{University College London}\\
London, United Kingdom \\}
\and
\IEEEauthorblockN{Christos Masouros}
\IEEEauthorblockA{\textit{Department of Electronic and Electrical Engineering} \\
\textit{University College London}\\
London, United Kingdom \\}}

% The paper headers
\markboth{Journal of \LaTeX\ Class Files,~Vol.~14, No.~8, August~2015}%
{Shell \MakeLowercase{\textit{et al.}}: Bare Demo of IEEEtran.cls for IEEE Journals}

\maketitle

\begin{abstract}
This study proposes an index modulation (IM) technique for affine frequency division multiplexing (AFDM) signals and examines its communication and sensing performance toward integrated sensing and communication (ISAC) systems. The power levels of subcarriers are utilized as modulation indices while also transmitting data symbols within each subcarrier. Thus, the proposed AFDM with
subcarrier power-level index modulation (AFDM-PLIM) maintains all subcarriers active at all times to achieve a higher spectral efficiency compared to other AFDM-IM techniques, where some of the subcarriers are turned off for IM. A low complexity estimator and subcarrier grouping are also proposed to reduce the computational complexity of the maximum likelihood estimator. Furthermore, this study also examines the delay and Doppler ambiguity functions of the proposed AFDM-PLIM and evaluates its range estimation performance. The results show that its sensing performance is better than AFDM-IM waveforms due to keeping all subcarriers active at all times.
\end{abstract}

\begin{IEEEkeywords}
Affine frequency division multiplexing, AFDM, Integrated sensing and communications, ISAC, index modulation.
\end{IEEEkeywords}

\IEEEpeerreviewmaketitle

\section{Introduction}

\IEEEPARstart{I}{ntegrated} sensing and communication (ISAC) is expected to be an essential part of the future generation of communication networks since many future applications such as autonomous driving or smart devices will demand high data rates and seamless wireless connectivity as well as highly precise and reliable sensing capabilities \cite{liu2022integrated}. Accordingly, ISAC has been one of the main research topics in recent years, from theoretical studies to waveform design and experimental measurements \cite{liu2022survey, temiz2021dual, TemizRadarCentric2023}.  Various waveforms have already been considered for future ISAC systems, such as orthogonal frequency division multiplexing (OFDM), orthogonal time frequency space (OTFS), AFDM, and frequency-modulated continuous wave (FMCW) signals \cite{TemizOptimized2021, RouOTFS_AFDM2024, TemizRadarCentrixConf2023}.

OTFS and AFDM signals are especially appealing for high-velocity scenarios due to their robustness to the intercarrier interference caused by the Doppler shifts \cite{bemani2023affine, RouOTFS_AFDM2024}. OTFS waveform utilizes the delay-Doppler domain to transmit data, while the AFDM waveform utilizes chirp-like subcarriers in the time-frequency domain similar to OFDM. Thus, both OTFS and AFDM waveforms can be used to communicate in linear time-varying (LTV) channels that have both high Doppler and delay variations, making them favorable in high-velocity scenarios, such as high-speed railway applications. The main benefits of AFDM compared to OTFS are lower pilot overheads and requiring lower complexity transmitter/receiver architectures \cite{tao2025affine, ranasinghe2024joint}.

Various index modulation-based improvements have already been proposed for OFDM and AFDM systems \cite{zhu2023design, liu2024pre, AnoopDualModeIMAFDM2025, MasourosDualLayerMIMO2016}. AFDM with on-off subcarrier index modulation is proposed in \cite{zhu2023design}, where the subcarriers are chosen to be off (disabled) or on (active) to carry IM data while active subcarriers also carry communication symbols.  This improves the BER at the expense of reducing the data rate since only active subcarriers can be used to transmit data via quadrature amplitude modulation (QAM) or phase-shift keying (PSK) symbols. Another study considered pre-chirp AFDM parameters to modulate IM data along with communication symbols \cite{liu2024pre}. Groups of subcarriers are also considered as indices of IM to transmit additional data \cite{AnoopDualModeIMAFDM2025}. In these studies, AFDM with index modulation has been shown to offer a better BER and energy-efficiency performance compared to AFDM. 

AFDM has also been considered for ISAC to sense the environment while performing communication with users \cite{ni2025integrated,zhu2024afdm,bemani2024integrated,ni2022afdm, BaoAFDMISAC2024}. An efficient target parameter estimation method that utilizes decoupled delay and Doppler in the chirp domain in \cite{ni2025integrated}. Another study investigates bistatic sensing via AFDM signals and proposes a bistatic sensing-aided channel estimation \cite{zhu2024afdm}. Moreover, the performance trade-off between communication and sensing based on AFDM parameters is investigated in \cite{BaoAFDMISAC2024}. These studies demonstrate that AFDM signals offer flexibility to design and realize ISAC systems.

In this study, we have considered AFDM with subcarrier power-level index modulation (AFDM-PLIM) for ISAC, where the power level of the subcarriers is utilized as IM indices. Different from previous AFDM-IM studies, all subcarriers are active and used to transmit symbols at all times in our approach; hence, it can achieve a higher communication sum rate compared to previous AFDM-IM approaches. Moreover, it has a better ambiguity function for sensing since all subcarriers are active at all times.

\section{System Model}

A single-cell ISAC setup is considered, where the base station (BS) is equipped with transmit antennas for ISAC signaling and to receive communication data from users (UEs), as well as radar receive antennas. In the rest of the article, we focus on a single-input single-output (SISO) mode for the sake of simplicity, and the signal and channel models are given accordingly\footnote{The system model and the proposed AFDM-PLIM can be extended to a multiple-input multiple-output (MIMO) scenario, and this will be the focus of the next study.}.

\subsection{AFDM with Subcarrier Power Level Index Modulation}
$\mathcal{M}$-PSK  modulated symbol data vector, that is intended to be transmitted to the user, is denoted by $\mathbf{s} \in \mathbb{C}^{L\times 1}$. It consists of PSK symbols for $L$ subcarriers, and IM bits are modulated using the power level of each subcarrier, hence,
\begin{equation}
    \mathbf{x}= \boldsymbol{\alpha} \mathbf{s} = \left[\sqrt{\alpha_1} s_1, \, \sqrt{\alpha_2} s_2,\, \dots, \, \sqrt{\alpha_L} s_L \right], 
\end{equation}
where $s_l$ denotes the modulated symbol, which is drawn from a $\mathcal{M}$-PSK  modulation constellation, and $\sqrt{\alpha_l}$ denotes the amplitude of the subcarrier, which corresponds to $z_l$ entry of IM data vector $\mathbf{z}$ such that
\begin{equation}
    \alpha_l =
\begin{cases}
    1+\beta,& \text{for }  z_l=1 \\
    1-\beta,& \text{for }  z_l=0. 
\end{cases}
\end{equation}
Modulating with IM data, as explained above, does not change the average power of the transmitted signals since the probability of having 1s and 0s are equal in the IM data such that $\mathcal{P}(z_l = 0) = \mathcal{P}(z_l = 1) = 0.5$. Moreover, the power of PSK symbols are the same, i.e., $\mathbb{E}\left[\left|s_l\right|^2\right]=1$ hence $\mathbb{E}\left[\left|\sqrt{\alpha_l}s_l\right|^2\right] = \mathbb{E}\left[\left|\sqrt{\alpha_l}\right|^2\right]\left[\left|s_l\right|^2\right] =\frac{1+\beta}{2}+\frac{1-\beta}{2}=1$ since $s_l$ and $\alpha_l$ are independent of each other.  Note that the choice of $0<\beta<1$ affects the demodulation performance of PSK and IM symbols since a larger $\beta$ improves the demodulation of IM but deteriorates the demodulation of PSK symbols. % by reducing the power of some symbols, i.e., $z_l=0$ case. 

\subsection{Affine Frequency Division Multiplexing with Subcarrier Power-Level Index Modulation}

AFDM and subcarrier power-level IM  (AFDM-PLIM) modulated signal in the time domain for the $k$th UE is given by
\begin{equation}
    {\mathbf y} = \mathbf{A}^H\mathbf{x}={\mathbf \Lambda}_{c_1}^H{\mathbf F}^H{\mathbf \Lambda}_{c_2}^H\mathbf{x},
\end{equation}
where data vector $\mathbf{x}$ is defined above, $\mathbf{A}=\mathbf{ \Lambda}_{c_1}{\mathbf F}{\mathbf \Lambda}_{c_2}$ denotes the discrete affine Fourier transform (DAFT) matrix, and  matrix $\mathbf{F}$ denotes the L-point discrete Fourier transform (DFT) matrix given by
\begin{equation}
    \mathbf{F}({k,l}) = e^{-j2\pi kl/L} \quad\quad k,l\in \{0,1,\dots,L-1\},
\end{equation}
and the diagonal matrices $\mathbf{\Lambda}_{c_1}$ and $\mathbf{\Lambda}_{c_2}$ are given by
\begin{equation}
    \mathbf{\Lambda}_{c_i} = \operatorname{diag} ([1, e^{-j2\pi c_1},\dots, e^{-j2\pi c_1(L-1)^2} ]),
\end{equation}
where $c_i\in\{c_1, c_2\}$. The DAFT parameters $c_1$ and $c_2$ are chosen as \cite{RouOTFS_AFDM2024},
\begin{equation}
    c_1 = \frac{2(\nu_{max}+\xi)+1}{2N}  \quad\quad c_2 << \frac{1}{2N},
\end{equation}
where $\nu_{max}$ denotes the channel's maximum normalized digital Doppler shift while $\xi$ is a small number, e.g., $\xi<5$ to include an additional guard against the Doppler shift. Parameter $c_2$ can be an arbitrarily small number satisfying the inequality above. As a result, the $l$th symbol transmitted in the time domain can be given by 
\begin{equation}
    \label{eq:mod}
    X_l = \frac{1}{\sqrt{L}}\sum_{l = 0}^{L-1}x[l]\cdot e^{j2\pi (c_1l^2 + c_2l^2 + kl/L)}.
\end{equation}

\subsection{Doubly Dispersive Channel Model}

Doubly dispersive channel models have effects of both delay and Doppler of the propagation paths; hence, their channel state information (CSI) matrices are not diagonal, unlike flat-fading channels. The doubly dispersive channel matrix $\mathbf{H}\in\mathbb{C}^{L\times L}$ for $L$ subcarriers is given below,
\begin{equation}
\mathbf{H} =
\begin{bmatrix}
h_{0,0} & h_{0,1} & \cdots & h_{0,L-1} \\
h_{1,0} & h_{1,1} & \cdots & h_{1,L-1} \\
\vdots & \vdots & \ddots & \vdots \\
h_{L-1,0} & h_{L-1,1} & \cdots & h_{L-1,L-1}
\end{bmatrix},    
\end{equation}
of which each entry $h_{k,l}$ is modeled as \cite{KeDoublyDispersive2004},
\begin{equation}
   h_{k,l} = \sum_{q=0}^{Q-1} \alpha_q \delta(k - \tau_q) e^{j 2\pi \nu_q k t},
\end{equation}
where $Q$ denotes the number of paths and sampling interval, respectively. Path-related parameters $\alpha_q$, $\nu_q$, $\tau_q$ denote the gain of the $q$th path, the Doppler shift of $q$th path, the delay of $q$th path, respectively.

\subsection{Receiver Processing}
The time-domain baseband signal received at the receiver is given by
\begin{equation}
    {\mathbf r} = {\mathbf H}\mathbf{y} + \mathbf{n},
\end{equation}
which is firstly processed by DAFT as
\begin{equation}
    \mathbf{A}{\mathbf r} = \mathbf{A}\left({\mathbf H}\mathbf{y} + \mathbf{n}\right),
\end{equation} and then channel equalization is performed via the  linear minimum mean-square error (LMMSE) equalizer to obtain the estimated symbol vector at the receiver as
\begin{equation}
    \hat{\mathbf{x}} = \underbrace{\left( \widetilde{\mathbf{H}}^{{H}} \widetilde{\mathbf{H}} + \sigma^2 \mathbf{I}_N \right)^{-1} \widetilde{\mathbf{H}}^{{H}}}_{\text{LMMSE Equalizer}} \underbrace{{\mathbf \Lambda}_{c_1}{\mathbf F}{\mathbf \Lambda}_{c_2}}_{\text{DAFT}}\left({\mathbf H}\mathbf{y} + \mathbf{n}\right),
\end{equation}
where $\widetilde{\mathbf{H}} = {\mathbf \Lambda}_{c_1}{\mathbf F}{\mathbf \Lambda}_{c_2}\mathbf{H}{\mathbf \Lambda}_{c_1}^H{\mathbf F}^H{\mathbf \Lambda}_{c_2}^H$ denotes the effective channel matrix \cite{bemani2022low}. Subsequently, the estimations of the PSK symbols and IM data are performed as explained above. 

\subsection{Symbol Estimation}

The optimal estimation of IM and $\mathcal{M}$-PSK data can be performed via the maximum likelihood estimator as,
\begin{equation}
    (\hat{\mathbf{z}}, \hat{\mathbf{s}}) = \arg \min_{{\mathbf{z}\in\{0, 1\}}, \mathbf{s} \in \mathcal{S}^K} \left\| \mathcal{J} - \hat{\mathbf{x}} \right\|^2,
\end{equation}
where $\mathcal{J}$ is the set of possible IM and PSK symbol combinations in $L$ subcarriers. Due to the large size of $\mathcal{J}$, the computational complexity of the maximum-likelihood (ML) estimator can be extremely high, i.e., $\mathcal{O}(2^L)$. Hence, the following low-complexity estimator has been proposed. The mean power of the symbols received in $L$ subcarriers is calculated as,
\begin{equation}
    \hat{p}_L= \frac{1}{L} \sum_{n=0}^{L-1} \left| \hat{x}_l \right|^2,
\end{equation}
and then the power of each symbol is compared to this threshold for the decision as
,\begin{equation}
    \hat{z}_l =
\begin{cases}
    1,& \text{for }  p_l \geq \hat{p}_L \\
    0,& \text{for }  p_l \leq \hat{p}_L, 
\end{cases}
\end{equation}
where $p_l = \left| \hat{x}_l \right|^2$. Subsequently, the normalized  $\mathcal{M}$-PSK symbols are obtained as
\begin{equation}
    \hat{s}_l =  \begin{cases}
    \dfrac{\hat{x}_l}{\sqrt{1+\beta}},& \text{for }  \hat{z}_l = 1 \\
    \dfrac{\hat{x}_l}{\sqrt{1-\beta}},& \text{for }  \hat{z}_0 = 1, 
\end{cases}
\end{equation}
and then these $\mathcal{M}$-PSK are demodulated.

\section{Communication Data Rate}

Considering that each subcarrier can carry an independent IM data bit, the maximum data rate of the AFDM-PLIM is given by, 
\begin{equation}
    R_{\gamma}=L \log_2 \mathcal{M} + L.
\end{equation}

The power level of the subcarriers can be determined independent of the $\mathcal{M}$-PSK symbols in the proposed IM modulation scheme; hence, this decouples the demodulation of IM and $\mathcal{M}$-PSK symbols. This enables IM bits to be obtained separately from the PSK bits, hence improving the demodulation performance. Moreover, grouping subcarriers into blocks for IM will improve the bit error rate (BER) performance at the expense of a slightly reduced data rate, as explained below.

$L$ subcarriers are divided into $G$ blocks, such that the block size of each block is given by $U=L/G$. Moreover, in each block, the $U/2$ subcarriers are set high, and the $U/2$ subcarriers are set low to have a uniform power distribution across the subcarriers within the block. In this case, the data rate that can be achieved is given by
\begin{equation}\label{eq:date_rate}
    R_{\gamma} = L\log_2\mathcal{M}+ G\log_2\binom U {U/2},
\end{equation}
where $U\geq 2$ and the combination term can be approximated as
\begin{equation}
\binom U {U/2} = \frac{U!}{(U/2)! (U/2)!} \approx \frac{2^U}{\sqrt{\pi U / 2}},
\end{equation}
using Sterling's approximation given by 
a! $\approx \sqrt{2 \pi a} \left(\frac{a}{e}\right)^{a}$, which provides a decent approximation when $a>4$, the achievable data rate of the proposed AFDM-PLIM is approximated as,
\begin{equation}
    R_{\gamma} \approx L \log_2 \mathcal{M} + L - \frac{G}{2} \log_2\left( \frac{\pi U}{2} \right),
\end{equation}
when $U>4$.

On the other hand, the maximum data rate that can be achieved with the traditional on-off IM, AFDM-IM, where only active subcarriers carry $\mathcal{M}$-PSK data, is given by,
\begin{equation}
    R_{\beta} = Z\log_2\mathcal{M} + \log_2\binom L {Z},
\end{equation}
where $Z$ denotes the number of active subcarriers, and is generally chosen as $Z\leq \frac{L}{2}$, while the maximum data rate of only $\mathcal{M}$-PSK modulated AFDM is given by 
\begin{equation}
R_{\alpha}=L\log_2\mathcal{M}.    
\end{equation}
Evaluating the data rates from these equations, it is evident that $R_{\beta} < R_{\alpha} < R_{\gamma}$, showing the higher data rate achieved by the proposed AFDM-PLIM.

\section{Sensing Performance}
The sensing performance of the proposed waveform is evaluated using the ambiguity function for delay (range) and Doppler (velocity) estimations of the waveforms. We have considered FMCW chirp as a sensing performance benchmark since it is a widely used sensing waveform due to its superior delay and Doppler estimation performance and simplicity \cite{sturm2011waveform}. Moreover, it can also be employed to design radar-centric ISAC systems \cite{TemizRadarCentrixConf2023}.

\subsection{Ambiguity Function}
The ambiguity function gives an insight into the response of a radar waveform to a time delay \( \tau \) and Doppler frequency shift \( \nu \). The ambiguity function of signal $x$ is given by \cite{stein1981algorithms},
\begin{equation}
    \psi_x(\tau, \nu) = \int_{-\infty}^{\infty} x(t) x^*(t - \tau) e^{j 2 \pi \nu t} dt,
\end{equation}
where \( x(t) \), \( \tau \), and \( \nu \) denote the transmitted signal, the time delay (related to the range), and the Doppler shift (related to velocity), respectively. Moreover, \( x^*(t) \) represents the complex conjugate of \( x(t) \). %The range ambiguity function takes only the effect of time delay into account and is given by,
%\begin{equation}
%    \psi_x(\tau) = \int_{-\infty}^{\infty} x(t) x^*(t - \tau) dt,
%\end{equation}
%which characterizes the ability to resolve targets' range for signal $x(t)$. The velocity ambiguity function considers only the effect of the Doppler frequency shift and is given by,
%\begin{equation}
%    \psi_x(\nu) = \int_{-\infty}^{\infty} x(t) e^{j 2 \pi \nu t} dt,
%\end{equation}
%which determines the system’s ability to determine targets' velocities. 

\subsection{Target Range Estimation Accuracy}
This study also evaluates the range estimation performance of the proposed AFDM-PLIM waveform in comparison with FMCW, AFDM, and AFDM-IM waveforms. For this aim, a monostatic sensing scenario consisting of a single target is simulated, where the range of the target is $T$ m from the ISAC transmitter and receiver. The normalized mean absolute error (NMAE) of range estimations are evaluated as,  
\begin{equation}
    \xi_T= \frac{1}{N\cdot T_i} \cdot{\sum_{i=1}^{N} \left| \hat{T}_i - T_i \right|},
\end{equation}
where $\hat{T}$ denotes the estimated range of the target by the ISAC, and $N$ is the number of scenarios evaluated. This NMAE is utilized as the range sensing accuracy performance metric to evaluate the range estimation performance of AFDM-IM.

\section{Numerical Results}

The communication and sensing performances of the proposed waveform in comparison with others are evaluated via simulations. In these simulations, a 2.4 GHz carrier frequency is considered, and 128 subcarriers are considered.    

\begin{figure}
    \centering
\includegraphics[width=1\linewidth]{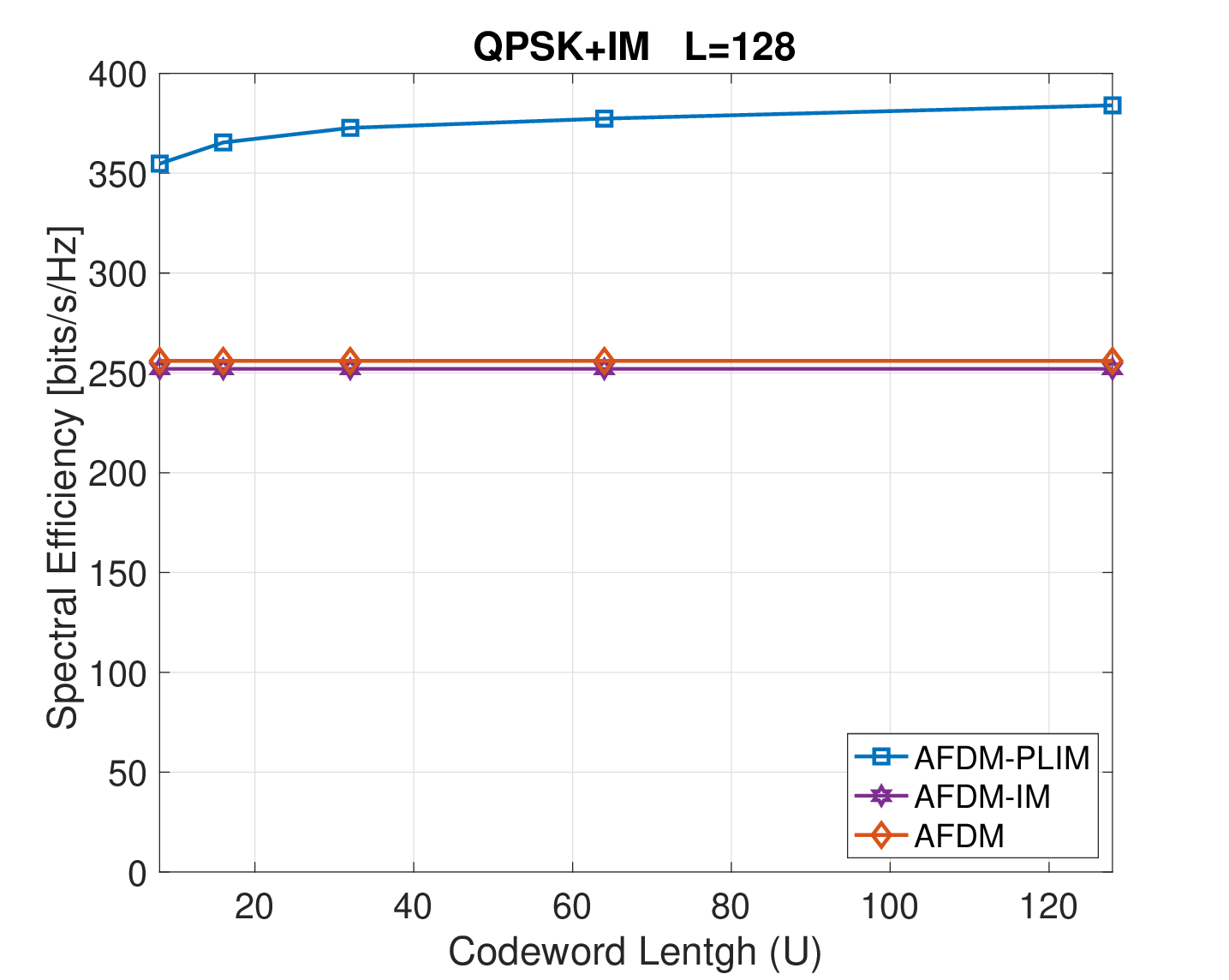}
    \caption{The spectral efficiency achieved by the proposed AFDM-PLIM in comparison with  AFDM-IM and AFDM. L=128 subcarriers and $Z=L/2$ for AFDM-IM.}
    \label{fig:spectral_effy}
\end{figure}

Fig.~\ref{fig:spectral_effy} illustrates the spectral efficiencies achieved by AFDM, AFDM-IM, and AFDM-PLIM. It is evident that even with short code-block length for IM, AFDM-PLIM has a higher spectral efficiency due to utilizing all subcarriers for $\mathcal{M}$-PSK symbols and IM data transmission. Moreover, choosing a shorter block size leads to a lower computational complexity for the ML estimator since its complexity can be given by $\mathcal{O}(2^U)$ for IM data estimation.

\begin{figure}
    \centering
\includegraphics[width=1\linewidth]{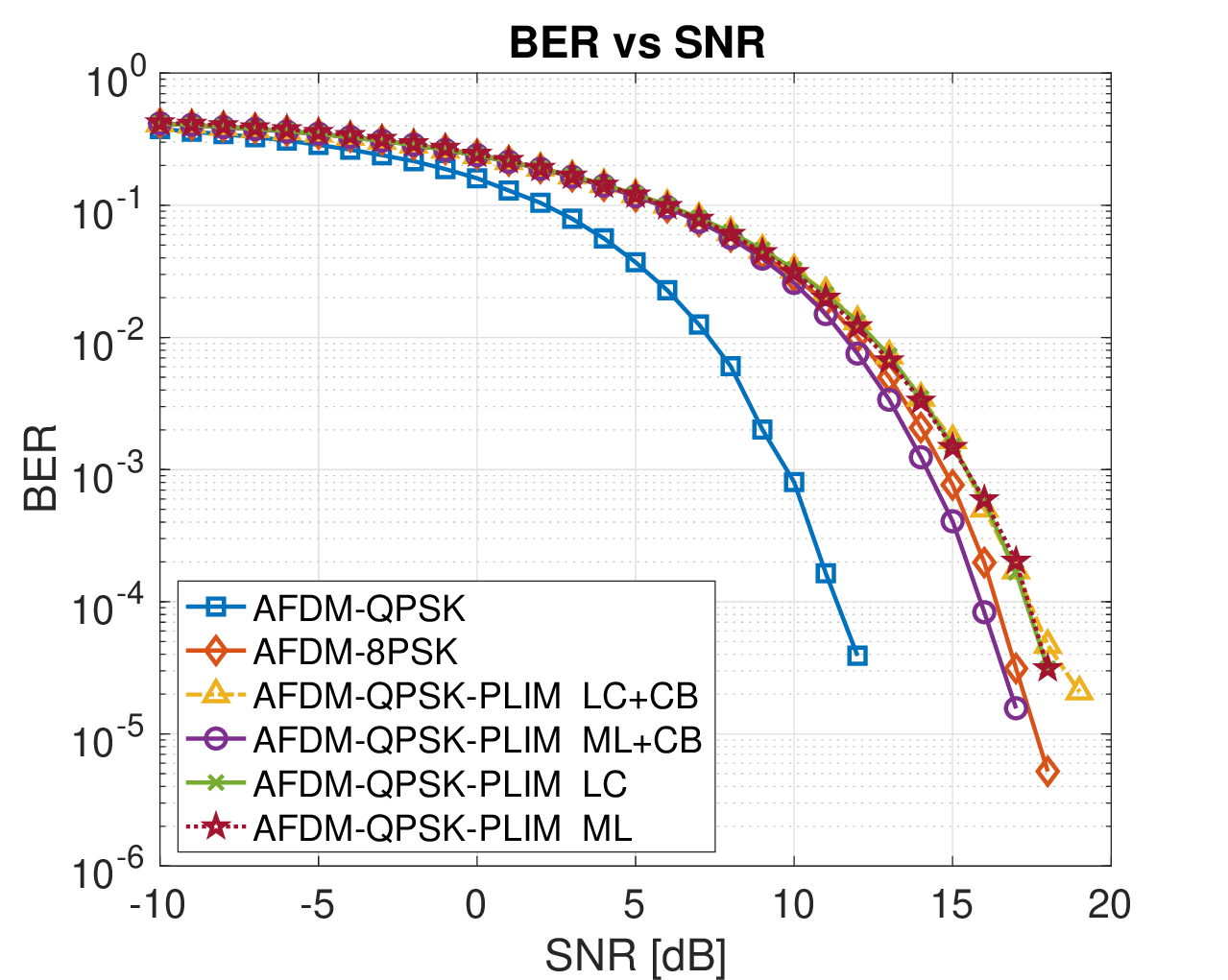}    \caption{The BER of the proposed AFDM-PLIM in comparison with AFDM-IM and AFDM.}
    \label{fig:BER}
\end{figure}

Fig.~\ref{fig:BER} presents the BER of the demodulations of the AFDM signal, where QPSK or 8PSK are employed in each subcarrier\footnote{Note that higher order PSK modulations can also be employed to achieve higher data rates}. This figure shows that the BER of the proposed low-complexity estimator (LC) is similar to the ML estimator. Moreover, when a code-block (CB) with U=64 is used, the proposed approach's BER performance is better than 8PSK while providing a similar data rate. By changing the block size, it is possible to achieve a better BER performance at the expense of slightly reduced data rate as shown in Fig.~\ref{fig:spectral_effy}. 

\begin{figure}
    \centering
\includegraphics[width=1\linewidth]{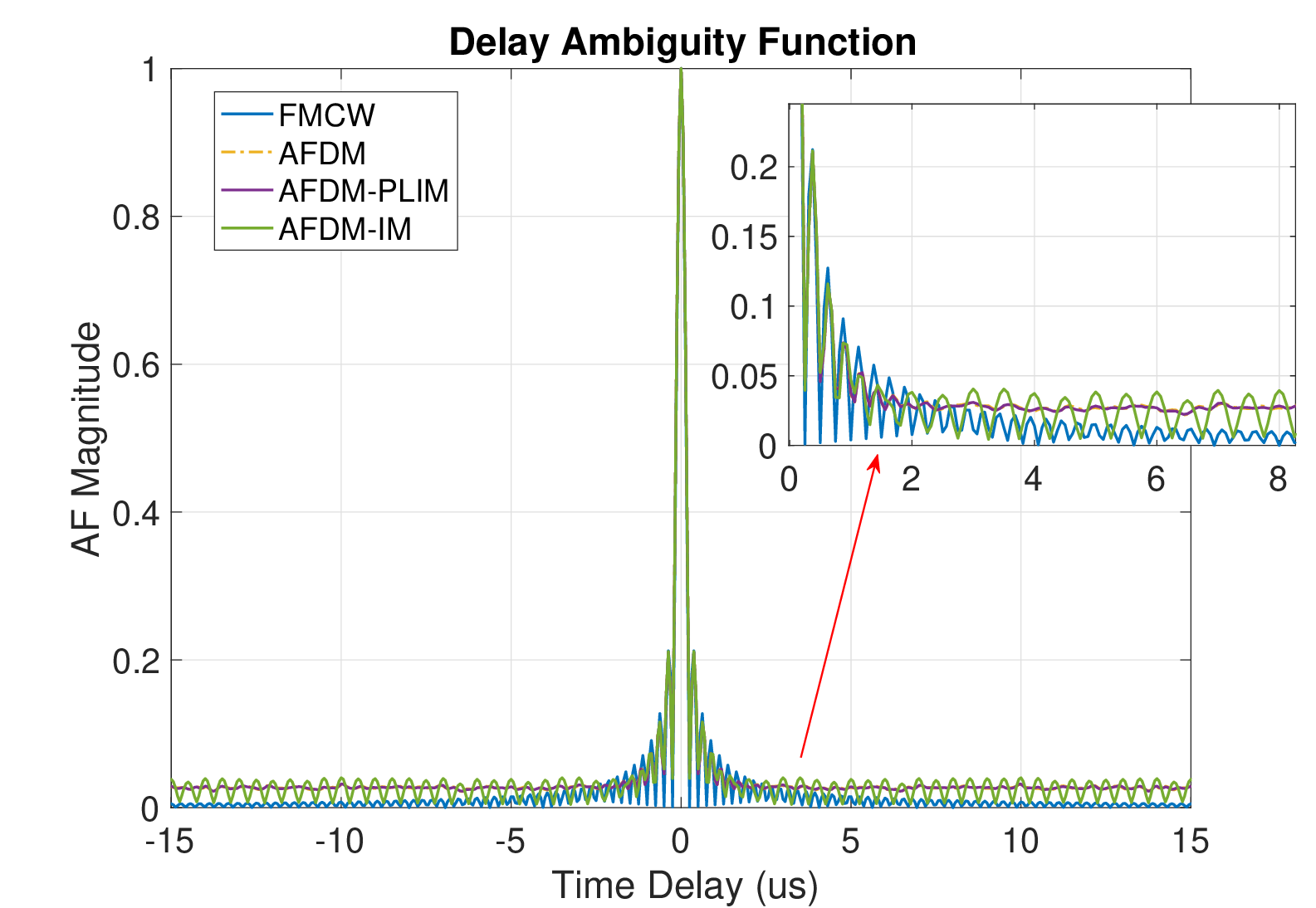}    \caption{The delay AF of the proposed AFDM-PLIM in comparison with AFDM-IM, AFDM, and FMCW waveforms.}
    \label{fig:Delay_AF}
\end{figure}

\begin{figure}
    \centering
\includegraphics[width=1\linewidth]{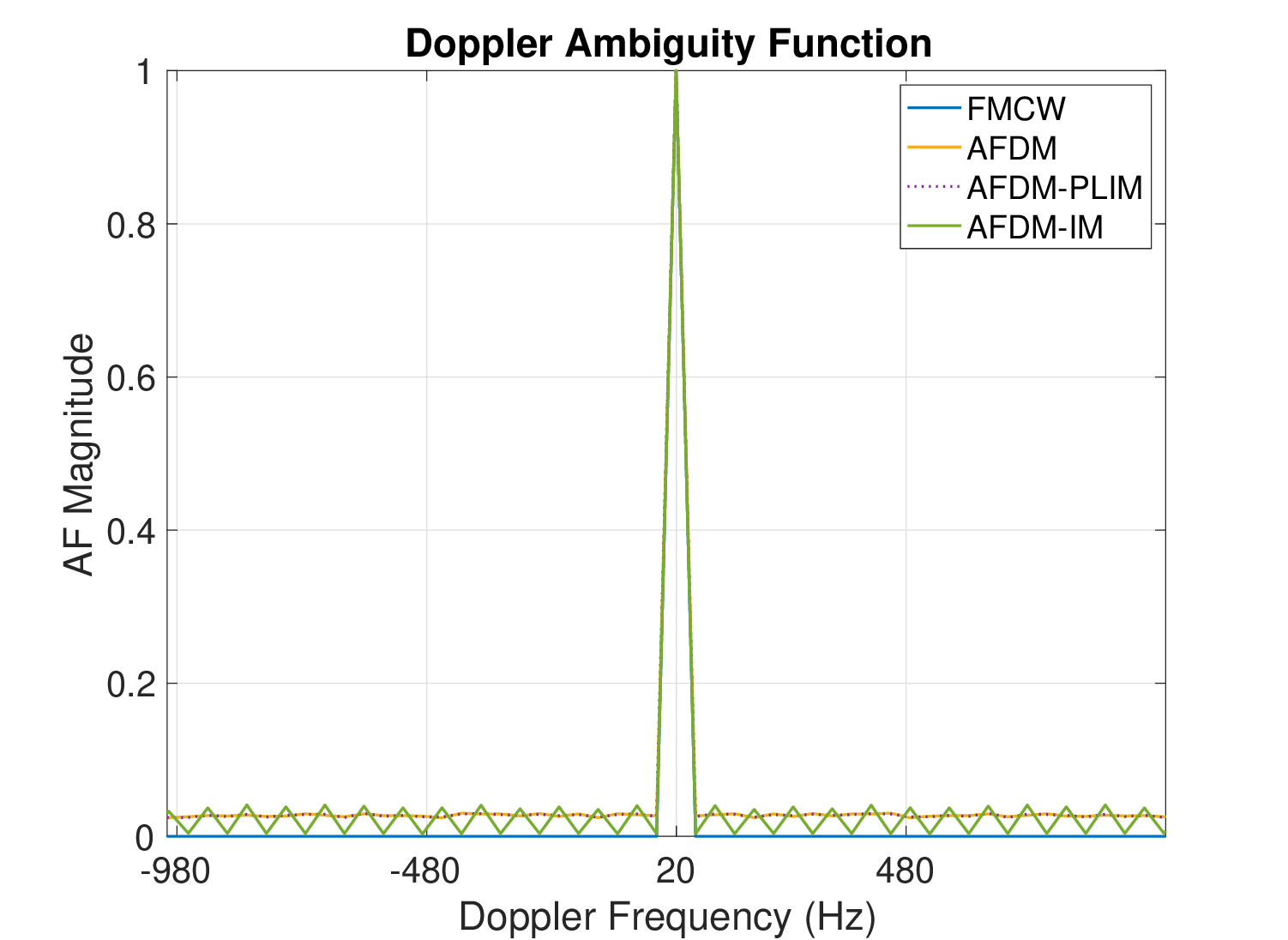}    \caption{The Doppler AF of the proposed AFDM-PLIM in comparison with AFDM-IM, AFDM, and FMCW waveforms.}
    \label{fig:Doppler_AF}
\end{figure}

Fig.~\ref{fig:Delay_AF} and Fig.~\ref{fig:Doppler_AF} show the delay and Doppler ambiguity functions of FMCW, AFDM, AFDM-IM, and AFDM-PLIM signals. It is clearly seen in Fig.~\ref{fig:Delay_AF} that AFDM-IM has oscillating sidelobes due to turning on and off subcarriers, while this issue is not seen in AFDM-PLIM. As expected, FMCW has the best delay and Doppler ambiguity functions, while the ambiguity function of the proposed AFDM-PLIM is similar to the AFDM waveform, and both have lower sidelobes than AFDM-IM. 

\begin{figure}
    \centering
\includegraphics[width=1\linewidth]{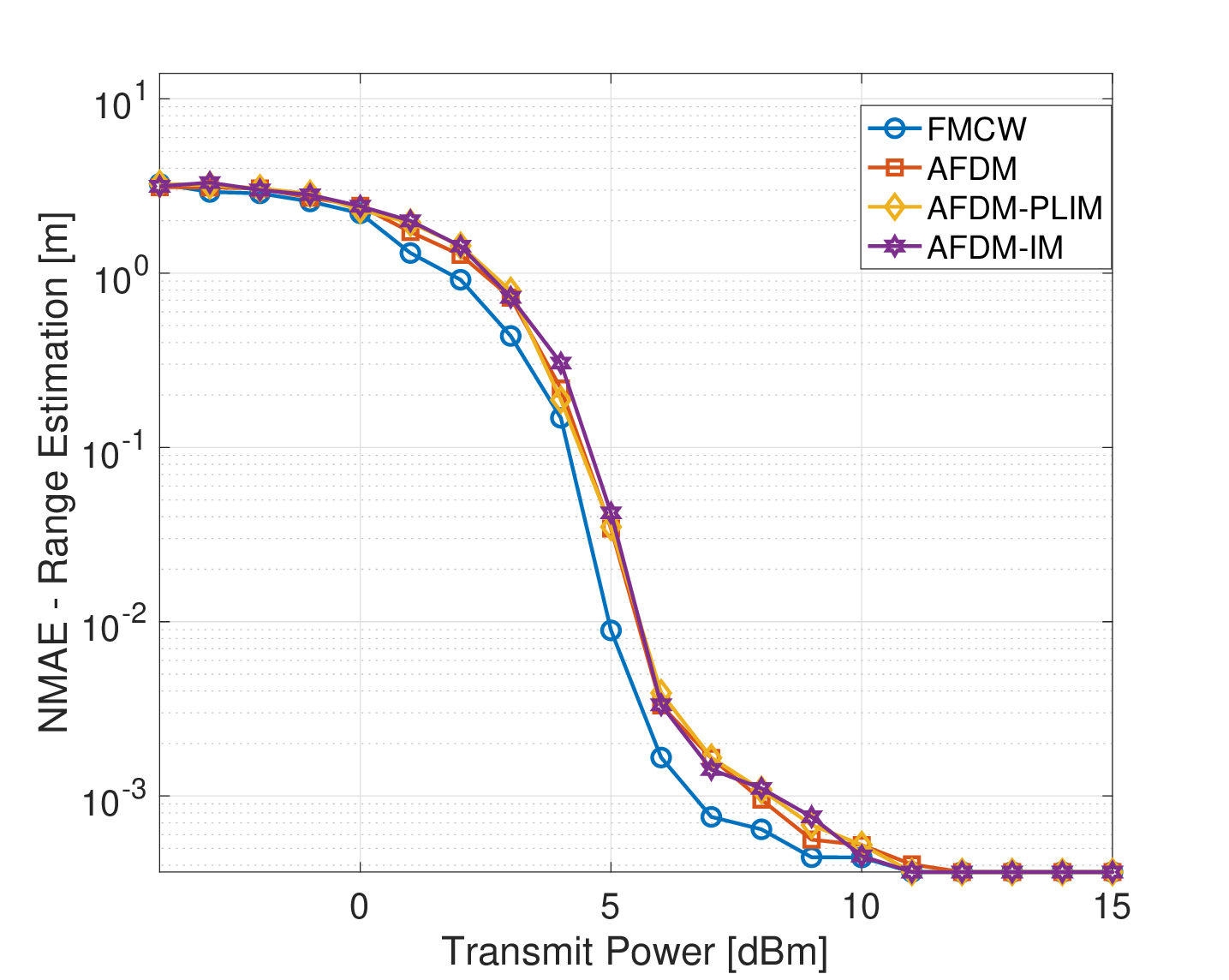}    \caption{The NMAE of range estimations obtained using AFDM-PLIM waveform in comparison with AFDM-IM, AFDM, and FMCW waveforms.}
    \label{fig:Range_NMSE}
\end{figure}

Fig.~\ref{fig:Range_NMSE} illustrates the range estimation accuracy of these waveforms in terms of the NMAE of range estimations as a function of transmit power. The target range $T=150$ m and $N=1000$ times target sensing simulations are repeated to obtain this figure. FMCW achieves the best range estimation performance due to its superior delay ambiguity function. The range estimation performance of AFDM-based waveforms is similar, however, AFDM and AFDM-PLIM perform slightly better than AFDM-IM according to these results. These results show that AFDM-PLIM is a good candidate for ISAC systems since it offers improved spectral efficiency and sensing performance than AFDM-IM.

\section{Conclusion}
This study has proposed the AFDM-PLIM waveform for integrated sensing and communications. The proposed waveform has been shown to achieve a higher spectral efficiency than AFDM and AFDM-IM, while it also offers better delay and Doppler ambiguity function characteristics than AFDM-IM. The computational complexity of the ML estimator can be extremely high for a large number of subcarriers; thus, we also proposed a low complexity AFDM-IM estimator. Moreover, a subcarrier grouping is proposed to improve its bit error rate (BER) performance at the expense of a slightly reduced data rate.

% use section* for acknowledgment
\section*{Acknowledgment}

This study has been funded by EPSRC Federated Telecoms Hub 6G Research Partnership Funds (THRPF) under the `Affine Frequency Division Multiplexing for 6G Communications and Sensing' project.

\ifCLASSOPTIONcaptionsoff
  \newpage
\fi

\bibliographystyle{IEEEtran}
\bibliography{ref.bib}

% Generated by IEEEtran.bst, version: 1.14 (2015/08/26)
\begin{thebibliography}{10}
\providecommand{\url}[1]{#1}
\csname url@samestyle\endcsname
\providecommand{\newblock}{\relax}
\providecommand{\bibinfo}[2]{#2}
\providecommand{\BIBentrySTDinterwordspacing}{\spaceskip=0pt\relax}
\providecommand{\BIBentryALTinterwordstretchfactor}{4}
\providecommand{\BIBentryALTinterwordspacing}{\spaceskip=\fontdimen2\font plus
\BIBentryALTinterwordstretchfactor\fontdimen3\font minus \fontdimen4\font\relax}
\providecommand{\BIBforeignlanguage}[2]{{%
\expandafter\ifx\csname l@#1\endcsname\relax
\typeout{** WARNING: IEEEtran.bst: No hyphenation pattern has been}%
\typeout{** loaded for the language `#1'. Using the pattern for}%
\typeout{** the default language instead.}%
\else
\language=\csname l@#1\endcsname
\fi
#2}}
\providecommand{\BIBdecl}{\relax}
\BIBdecl

\bibitem{liu2022integrated}
F.~Liu, Y.~Cui, C.~Masouros, J.~Xu, T.~X. Han, Y.~C. Eldar, and S.~Buzzi, ``Integrated sensing and communications: Toward dual-functional wireless networks for {6G} and beyond,'' \emph{IEEE journal on selected areas in communications}, vol.~40, no.~6, pp. 1728--1767, 2022.

\bibitem{liu2022survey}
A.~Liu, Z.~Huang, M.~Li, Y.~Wan, W.~Li, T.~X. Han, C.~Liu, R.~Du, D.~K.~P. Tan, J.~Lu \emph{et~al.}, ``A survey on fundamental limits of integrated sensing and communication,'' \emph{IEEE Communications Surveys \& Tutorials}, vol.~24, no.~2, pp. 994--1034, 2022.

\bibitem{temiz2021dual}
M.~Temiz, E.~Alsusa, and M.~W. Baidas, ``A dual-function massive {MIMO} uplink {OFDM} communication and radar architecture,'' \emph{IEEE Transactions on Cognitive Communications and Networking}, vol.~8, no.~2, pp. 750--762, 2021.

\bibitem{TemizRadarCentric2023}
M.~Temiz, C.~Horne, N.~J. Peters, M.~A. Ritchie, and C.~Masouros, ``An experimental study of radar-centric transmission for integrated sensing and communications,'' \emph{IEEE Transactions on Microwave Theory and Techniques}, vol.~71, no.~7, pp. 3203--3216, July 2023.

\bibitem{TemizOptimized2021}
M.~Temiz, E.~Alsusa, and M.~W. Baidas, ``Optimized precoders for massive {MIMO} {OFDM} dual radar-communication systems,'' \emph{IEEE Transactions on Communications}, vol.~69, no.~7, pp. 4781--4794, 2021.

\bibitem{RouOTFS_AFDM2024}
H.~S. Rou, G.~T.~F. de~Abreu, J.~Choi, D.~González~G., M.~Kountouris, Y.~L. Guan, and O.~Gonsa, ``From orthogonal time–frequency space to affine frequency-division multiplexing: A comparative study of next-generation waveforms for integrated sensing and communications in doubly dispersive channels,'' \emph{IEEE Signal Processing Magazine}, vol.~41, no.~5, pp. 71--86, Sep. 2024.

\bibitem{TemizRadarCentrixConf2023}
M.~Temiz, N.~J. Peters, C.~Horne, M.~A. Ritchie, and C.~Masouros, ``Radar-centric {ISAC} through index modulation: Over-the-air experimentation and trade-offs,'' in \emph{2023 IEEE Radar Conference (RadarConf23)}, 2023, pp. 1--6.

\bibitem{bemani2023affine}
A.~Bemani, N.~Ksairi, and M.~Kountouris, ``Affine frequency division multiplexing for next generation wireless communications,'' \emph{IEEE Transactions on Wireless Communications}, vol.~22, no.~11, pp. 8214--8229, 2023.

\bibitem{tao2025affine}
Y.~Tao, M.~Wen, Y.~Ge, J.~Li, E.~Basar, and N.~Al-Dhahir, ``Affine frequency division multiplexing with index modulation: Full diversity condition, performance analysis, and low-complexity detection,'' \emph{IEEE Journal on Selected Areas in Communications}, 2025.

\bibitem{ranasinghe2024joint}
K.~R.~R. Ranasinghe, H.~S. Rou, G.~T.~F. De~Abreu, T.~Takahashi, and K.~Ito, ``Joint channel, data and radar parameter estimation for {AFDM} systems in doubly-dispersive channels,'' \emph{IEEE Transactions on Wireless Communications}, 2024.

\bibitem{zhu2023design}
J.~Zhu, Q.~Luo, G.~Chen, P.~Xiao, and L.~Xiao, ``Design and performance analysis of index modulation empowered {AFDM} system,'' \emph{IEEE wireless communications letters}, vol.~13, no.~3, pp. 686--690, 2023.

\bibitem{liu2024pre}
G.~Liu, T.~Mao, R.~Liu, and Z.~Xiao, ``Pre-chirp-domain index modulation for affine frequency division multiplexing,'' in \emph{2024 International Wireless Communications and Mobile Computing (IWCMC)}.\hskip 1em plus 0.5em minus 0.4em\relax IEEE, 2024, pp. 0473--0478.

\bibitem{AnoopDualModeIMAFDM2025}
A.~Anoop, C.~K. Thomas, S.~Kala, J.~B. Benifa, and W.~Saad, ``Dual-mode index modulation based on affine frequency division multiplexing,'' \emph{Physical Communication}, vol.~70, p. 102628, 2025.

\bibitem{MasourosDualLayerMIMO2016}
C.~Masouros and L.~Hanzo, ``Dual-layered {MIMO} transmission for increased bandwidth efficiency,'' \emph{IEEE Transactions on Vehicular Technology}, vol.~65, no.~5, pp. 3139--3149, May 2016.

\bibitem{ni2025integrated}
Y.~Ni, P.~Yuan, Q.~Huang, F.~Liu, and Z.~Wang, ``An integrated sensing and communications system based on affine frequency division multiplexing,'' \emph{IEEE Transactions on Wireless Communications}, 2025.

\bibitem{zhu2024afdm}
J.~Zhu, Y.~Tang, F.~Liu, X.~Zhang, H.~Yin, and Y.~Zhou, ``{AFDM}-based bistatic integrated sensing and communication in static scatterer environments,'' \emph{IEEE Wireless Communications Letters}, 2024.

\bibitem{bemani2024integrated}
A.~Bemani, N.~Ksairi, and M.~Kountouris, ``Integrated sensing and communications with affine frequency division multiplexing,'' \emph{IEEE Wireless Communications Letters}, 2024.

\bibitem{ni2022afdm}
Y.~Ni, Z.~Wang, P.~Yuan, and Q.~Huang, ``An {AFDM}-based integrated sensing and communications,'' in \emph{2022 International Symposium on Wireless Communication Systems (ISWCS)}.\hskip 1em plus 0.5em minus 0.4em\relax IEEE, 2022, pp. 1--6.

\bibitem{BaoAFDMISAC2024}
H.~Bao, H.~Zhuang, Z.~Wang, and G.~Pang, ``Performance trade-off between communication and sensing based on {AFDM} parameter adjustment,'' in \emph{2024 IEEE 35th International Symposium on Personal, Indoor and Mobile Radio Communications (PIMRC)}, 2024, pp. 1--6.

\bibitem{KeDoublyDispersive2004}
K.~Liu, T.~Kadous, and A.~Sayeed, ``Orthogonal time-frequency signaling over doubly dispersive channels,'' \emph{IEEE Transactions on Information Theory}, vol.~50, no.~11, pp. 2583--2603, Nov 2004.

\bibitem{bemani2022low}
A.~Bemani, N.~Ksairi, and M.~Kountouris, ``Low complexity equalization for {AFDM} in doubly dispersive channels,'' in \emph{ICASSP 2022-2022 IEEE International Conference on Acoustics, Speech and Signal Processing (ICASSP)}.\hskip 1em plus 0.5em minus 0.4em\relax IEEE, 2022, pp. 5273--5277.

\bibitem{sturm2011waveform}
C.~Sturm and W.~Wiesbeck, ``Waveform design and signal processing aspects for fusion of wireless communications and radar sensing,'' \emph{Proceedings of the IEEE}, vol.~99, no.~7, pp. 1236--1259, 2011.

\bibitem{stein1981algorithms}
S.~Stein, ``Algorithms for ambiguity function processing,'' \emph{IEEE Transactions on Acoustics, Speech, and Signal Processing}, vol.~29, no.~3, pp. 588--599, 1981.

\end{thebibliography}

\end{document}